\documentclass{article}
\usepackage[utf8]{inputenc}
\usepackage{graphicx}
\usepackage[dvipsnames]{xcolor}
\usepackage{footmisc}
\usepackage{caption}
\usepackage{subcaption}
\usepackage{chngpage}
\usepackage{authblk}
\usepackage{soul}
\usepackage{hyperref}

\mathchardef\mhyphen="2D
\title{Segmentation of Lung Tumor from CT Images using Deeply Supervised MultiResUNet\thanks{Farheen and Shamil contributed equally.}}

\title{Segmentation of Lung Tumor from CT Images using Deep Supervision\thanks{Farheen and Shamil contributed equally.}}

\author[1,2]{Farhanaz Farheen}
\author[1,2]{Md. Salman Shamil}
\author[1]{Nabil Ibtehaz}
\author[1]{M. Sohel Rahman \thanks{Corresponding Author: \url{msrahman@cse.buet.ac.bd}}}
\affil[1]{Department of CSE, BUET, ECE Building, West Palasi, Dhaka-1230, Bangladesh}
\affil[2]{Department of CSE, United International University, Dhaka, Bangladesh}
\date{}

\begin{document}

\maketitle

\section*{Abstract}
Lung cancer is a leading cause of death in most countries of the world. Since prompt diagnosis of tumors can allow oncologists to discern their nature, type and the mode of treatment, tumor detection and segmentation from CT Scan images is a crucial field of study worldwide. This paper approaches lung tumor segmentation by applying two-dimensional discrete wavelet transform (DWT) on the LOTUS dataset for more meticulous texture analysis whilst integrating information from neighboring CT slices before feeding them to a Deeply Supervised MultiResUNet model. Variations in learning rates, decay and optimization algorithms while training the network have led to different dice co-efficients, the detailed statistics of which have been included in this paper. We also discuss the challenges in this dataset and how we opted to overcome them. In essence, this study aims to maximize the success rate of predicting tumor regions from two-dimensional CT Scan slices by experimenting with a number of adequate networks, resulting in a dice co-efficient of 0.8472. \\
\textbf{Keywords:} Lung tumor, CT scan images, deep learning, discrete wavelet transform.

\section{Introduction}
Cancer is one of the most fatal afflictions that can affect the human body and is indisputably a leading cause of death worldwide. Lung cancer is the most commonly diagnosed cancer in the world with instances constituting 11.6\% of total cancer cases and is also the prime cause of cancer-related fatalities, covering 18.4\% of them \cite{bray2018global}. 
Apart from facing distressing chest pain, weight loss and cough, lung cancer patients are more likely to suffer from chronic obstructive pulmonary disease \cite{ettinger2010non}. Mutation of lung cells - often prompted by imprudent smoking, which is the primary cause of lung cancer or simply due to exposure to Radon gas \cite{ettinger2010non} - followed by their disorderly growth, results in formation of tumors. Lung tumors are mainly aberrant clusters of tissues that are outcomes of atypical cell division frequency in the lungs \cite{lungtumor1}, \cite{lungtumor2}. Diagnosis of malignant tumors is often not feasible in early stages for the lack of symptoms or because the symptoms are indistinguishable from those of a respiratory infection. Most lung cancer cases are thus diagnosed at an advanced stage \cite{midthun2016early}, inevitably leading to the patient’s death, which urgently calls for a coherent approach towards developing a method to deal with lung cancer. 

Since the symptoms of lung cancer appear and the cases are diagnosed at an advanced stage most of the time \cite{midthun2016early}, it is practically impossible to save the patient in the end. It is crucial to make an effort to detect it at the preliminary stages because early diagnosis can allow the patient to go through radiotherapy or chemotherapy when the tumors haven’t infected neighboring tissues yet. This is rather convenient for oncologists as they can determine how to approach the treatment of the patient. An important observation in this regard is that lung CT scan images have the capability of revealing tumors at an early stage. The challenge, however, is that these attempts are tedious, time consuming and more prone to errors in general \cite{makaju2018lung}, \cite{nadkarni2019detection}.

Lung cancer prediction pipeline using CT Scan images goes through several stages including image pre-processing, segmentation, feature extraction and classification. The pre-processing technique used in \cite{sangamithraa2016lung} involves removing unwanted artifacts with the help of Median and Wiener filters. Next, a K-means clustering method is applied for segmentation followed by an EK-Means clustering. In \cite{kaucha2017early}, an SVM classifier is used for early detection of lung cancer. After pre-processing the CT scan image of the lungs, the Region of Interest (ROI) is segmented, retained and compressed using Discrete Wavelet Transform. Another work that uses Support Vector Machines for lung cancer classification is reported in \cite{mithuna2018quantitative}. Patel et al. used the local energy-based shape histogram (LESH) feature extraction technique with sensitivity analysis (SA) to detect lung cancer \cite{patel2018hybrid}. In \cite{perumal2018lung}, an enhanced artificial bee colony optimization technique is used for lung cancer detection and classification. Most common procedures for lung cancer detection have had a significant reliance on typical Image processing, machine learning based on handcrafted features and soft computing techniques.

In recent times, deep learning techniques have revolutionized the field of medical image processing and analysis. Haque et al. in \cite{haque2020deep} have discussed the superiority of deep learning techniques for the purpose of medical image segmentation. Traditional machine learning approaches usually cannot process natural data in their raw form and are incapable of dynamically adapting to new information \cite{haque2020deep}. However, this is something that deep learning techniques are good at and so they are widely used for handling data in their raw form. Moreover, with the advent of advanced central processing units and graphic processing units, training and execution times have reduced as well, making it more suitable to use deep learning based algorithms. Therefore, there has been a recent emergence of advanced deep learning techniques \cite{lecun2015deep} including Convolutional Neural Networks (CNN) \cite{lecun2010convolutional} in the area of medical image segmentation. For example, in \cite{alakwaa2017lung}, lung cancer detection and classification is done using 3D Convolutional Neural Networks. The Kaggle Data Science Bowl 2017 (KDSB17) \cite{kaggle2017kaggle} challenge saw multitudinous applications of Convolutional Neural Networks on KDSB17 datasets accompanied by LUNA16 dataset 
in some cases \cite{kuan2017deep}. On the other hand, the 2018 VIP-CUP Challenge presented the problem of segmentation and prediction of Lung tumor regions that was addressed with methodical procedures like Recurrent 3D-DenseUNet architecture using the tversky loss function \cite{kamal2018lung}. Furthermore, Dilated hybrid 3D Convolutional Neural Networks have also been experimented with in this context apart from LungNet and UNet models \cite{hossain2019pipeline}. All these approaches reported excellent results. 

In \cite{xu2014deep}, a study of automatic extraction of features is presented leveraging deep learning techniques. A CAD system is proposed in \cite{kumar2015lung} that uses deep features extracted from an autoencoder to classify lung nodules as malignant or benign. Moreover, three deep learning approaches, namely, Convolutional Neural Network (CNN), Deep Belief Networks (DBNs) and Stacked Denoising Autoencoder (SDAE) are applied in \cite{sun2016computer} to demonstrate the use of deep learning in lung cancer diagnosis using the Lung Image Database Consortium (LIDC) database.


The main contribution of this paper is our newly developed pipeline that consists of a preprocessing phase with efficient feature engineering, an existing deep learning architecture with some useful enhancements and post-processing techniques for improvement in the results. 3D Convolutional Neural Networks, which are quite compute- and memory-intensive, may lead to overfitting with small datasets because of a larger number of parameters, which often calls for the alternative of conceptualizing the 3D space as a collection of 2D planes. To this end, we have converted the problem in such a way that we can use 2D kernels with our designed feature map and achieved much better performance while maintaining efficiency in terms of memory usage and runtime. Thus our technical contribution in this paper involves (a) adopting deep supervision \cite{lee2015deeply} with MultiResUnet framework - a considerable improvement from the state of the art models \cite{ibtehaz2020multiresunet}- after (b) applying two dimensional discrete wavelet transform on the dataset for inferring the texture better and (c) analyzing neighboring CT slices for more confident deduction.  
Additionally, the concerning bottleneck posed by a small dataset has been handled by an attempt to diversify it through data augmentation via rotations by differing angles. All of these phases and their combined impact on accurate segmentation have been the most important contribution of our project which ultimately resulted in a dice co-efficient of 0.8472 - a significant advancement over the prevalent assessments.

\section{Materials and Methods}\label{materials}
\subsection{Dataset}
Our experiments have been conducted on the LOTUS Dataset (supplied by MAASTRO clinic) - an amended adaptation of the NSCLC-Radiomics Dataset - consisting of images and annotations of 300 patients in total, i.e., from 260 patients for the training set and from 40 patients for the validation set \cite{lotusbenchmark}. However, we used the validation set as an independent test set since no labels were supplied to accompany the original test set. The annotations of this dataset cover a wide region surrounding the tumor area including both the right and left lungs. Tumor volumes - Gross Tumor Volume, Planning Tumor Volume and Clinical Target Volume - have also been annotated in this dataset \cite{lotusbenchmark}. It is a compact collection of DICOM files containing several medical details on each patient. Our preprocessing methods extract the ${512 \times 512}$ CT scan slices from these DICOM objects that are sent into the pipeline after some further partitioning and refinements. Table \ref{tab:my_label} provides a summary of the dataset \cite{lotusbenchmark}.

\begin{table}[!htbp]
    \centering
    \caption{Number of tumor and non-tumor slices in dataset}\label{tab:my_label}
    \begin{tabular}{|c|c|c|c|}
        \hline
         Dataset & Number of Subjects & Tumor Slices & Non-Tumor Slices\\
         \hline
         Training Set & 260 & 4296 & 26951\\
         Test Set & 40 & 848 & 3610\\
         \hline
    \end{tabular}
        
\end{table}

Another important detail to be noted is the fact that the CT Scanners differed for the slices from different patients. Two manufacturers, namely, CMS Imaging Inc. and SIEMENS were used for this purpose. The relevant statistics have been reported in Table \ref{tab:table2} \cite{lotusbenchmark}. 

\begin{table}[!htbp]
    \centering
    \caption{Dataset statistics with respect to different manufacturers}\label{tab:table2}
    \begin{tabular}{|c|c|c|c|}
        \hline
         Dataset & Number of Patients & SIEMENS & CMS Imaging Inc.\\
         \hline
         Training Set & 260 & 200 & 60\\
         Test Set & 40 & 6 & 34\\
         \hline
    \end{tabular}

\end{table}

The dataset contained folders having the slices and annotation for each patient. A few of these were dropped owing to missing or inconsistent annotations. Apart from those, all the other folders contained a number of DICOM files along with their corresponding contours. There were some minor inconsistencies involving the identifiers of the annotation files. They were handled by examining the co-ordinates of the original CT slice.

\subsection{Proposed Approach}
\subsubsection{Overview of the pipeline}

In our study, we have used two dimensional CT slices for our segmentation task. We have applied two dimensional discrete wavelet transforms on the original input images. We have also utilized the neighboring CT slices with each image during the input instance preparation. Next, for training the dataset, we have used UNet, MultiResUNet as well as deep supervision with both of these models. We have also used Test Time Augmentation after that.

Challenges in medical image segmentation include variation in the shape, size and texture of the Region of Interest (ROI) \cite{haque2020deep}. Our dataset contains a very small Region of Interest (ROI) in the two dimensional CT slices. If we had taken three dimensional slices, the proportion of non tumor pixels would have increased immensely. This would have significantly affected the performance of the deep learning models as the models would have become biased towards non-tumor regions. Moreover, the use of three dimensional slices would have required a huge number of parameters, making the task more computationally expensive. On the contrary, we have used two dimensional slices which demand less parameters. To ensure that we are involving the minimum spatial information in the input instance, we have added the neighboring CT slices with each original slice. However, unlike a three dimensional approach, it does not significantly inflate the proportion of blank space in the input, neither does it increase the parameter usage to a great extent.

Since the ROI is already quite small, capturing the texture of the tumors in the dataset is a rather arduous task and as mentioned before, texture variation in ROI can be a challenge. Wavelet transforms can perform this task of texture analysis very well \cite{livens1997wavelets}. It allows the model to identify minute details in the image because of which, we have chosen to use this in our preprocessing stage.

\subsubsection{Preprocessing}

Based on the manufacturer, the DICOM files were initially partitioned into two groups. There was a stark difference in their span of intensity values in Hounsfield Unit (HU): from -1024 to 3071 for CMS Imaging Inc. and from 0 to 4095 for SIEMENS. Each partition was normalized to establish relative uniformity in the dataset.

The masks generated had a lot of blank spaces surrounding the tumor region. This made the tumor fairly insignificant compared to the empty area. The images were, thus, cropped to a smaller window to ensure that an adequately amplified tumor portion is supplied to the model. We used the same co-ordinates to crop both the training and test sets, obtaining this window by analyzing the masks of the training set.

The preprocessing step also involved gathering more information from neighboring CT slices. A tumor is a three dimensional structure and these CT images are only two dimensional slices of them. Principally, this calls for a three dimensional processing scheme which, however, is memory intensive. Nevertheless, whether a two dimensional slice would contain a tumor region or not would depend on its adjacent slices to a great extent. Therefore, to bring appropriate context and to account for dependencies between consecutive slices, the neighbors were added to each slice for preparing the input instance.

A two-dimensional discrete wavelet transform was applied on each CT slice twice - generating first and second approximations of the image \cite{lee2019pywavelets}. This operation simplifies images by leaving out some sharp horizontal, vertical and diagonal details. By analyzing the original as well as the transformed slices, the model can perceive tumor textures more minutely by identifying where these details were lost. In our study, the ${512 \times 512}$ images were resized to ${128 \times 128}$.

\begin{figure}[!htbp]
     \centering
     \includegraphics[scale=0.41]{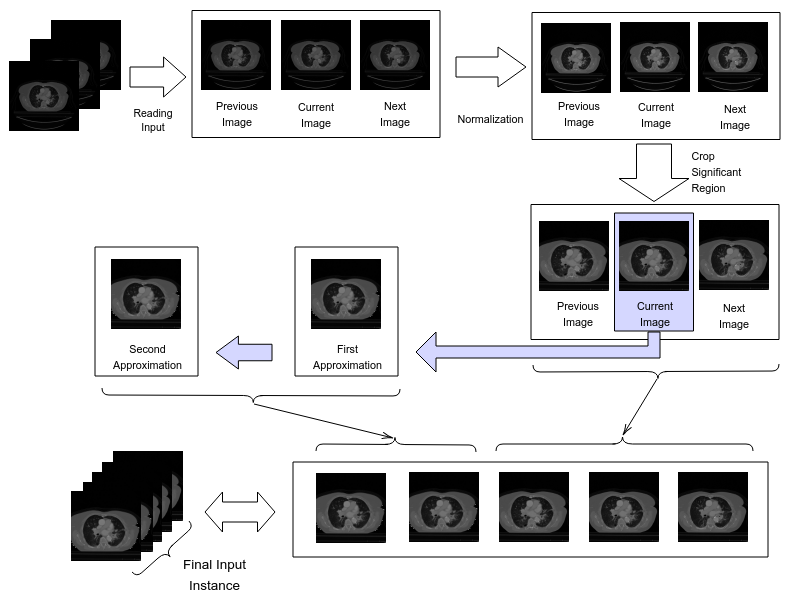}
     \caption{Input Instance Preparation}
     \label{fig:preprocess}
 \end{figure}

The final feature set comprised five slices concatenated one after another representing the original image, the previous and the next images, the first and the second approximations of the original image (by two-dimensional discrete wavelet transform). The training set was enlarged, or more accurately, doubled by augmenting the dataset, i.e., flipping or rotating each image based on a probability. Figure \ref{fig:preprocess} illustrates the pre-processing steps to prepare the input instances.

Finally, we performed Test Time Augmentation (TTA) by creating multiple augmented copies of each image in the test set. This was done by rotating each image between 20 to 100 times. 

\subsubsection{Description of the Models}
Since we experimented with two dimensional U-Net \cite{ronneberger2015u}, MultiResUNet \cite{ibtehaz2020multiresunet} and Deeply Supervised \cite{lee2015deeply} networks, the input instances of dimension 128 x 128 x 5 were fed into all the architectures separately. 
\begin{enumerate}
  \item \textbf{U-Net}\\
  The U-Net framework consists of an encoding phase and a decoding phase. Each layer covers two convolutions and a Max-Pooling operation in the encoding phase whereas the decoding phase replaces the Max-Pooling operation with an Up-sampling. As the data progresses down the encoders, it gradually secures more context while leaving out details about location. Consequently, spatial information is prepended to the contextual data of the decoders via skip connections.

  \item \textbf{MultiResUNet}\\
 The generic design of this network is quite similar to that of U-Net, except for slight differences in the nature of convolutions in each layer and the residual path from the encoders to the decoders (Figure \ref{fig:multiresunet}). Each layer of U-Net is replaced by a MultiRes block that contains 4 convolutions (Figure \ref{fig:mblock}). The skip connections of U-Net are replaced by Res Paths (Figure \ref{fig:respath}) as described in \cite{ibtehaz2020multiresunet}. The motivation of adopting this network comes from the idea that an ideal architecture should be able to assess images having diversified scales in medical image segmentation procedures. For further details readers are referred to \cite{ibtehaz2020multiresunet}. 
 
 

 \begin{figure}[!htbp]
     \centering
    \begin{subfigure}[b]{\textwidth}
         \includegraphics[width=\textwidth]{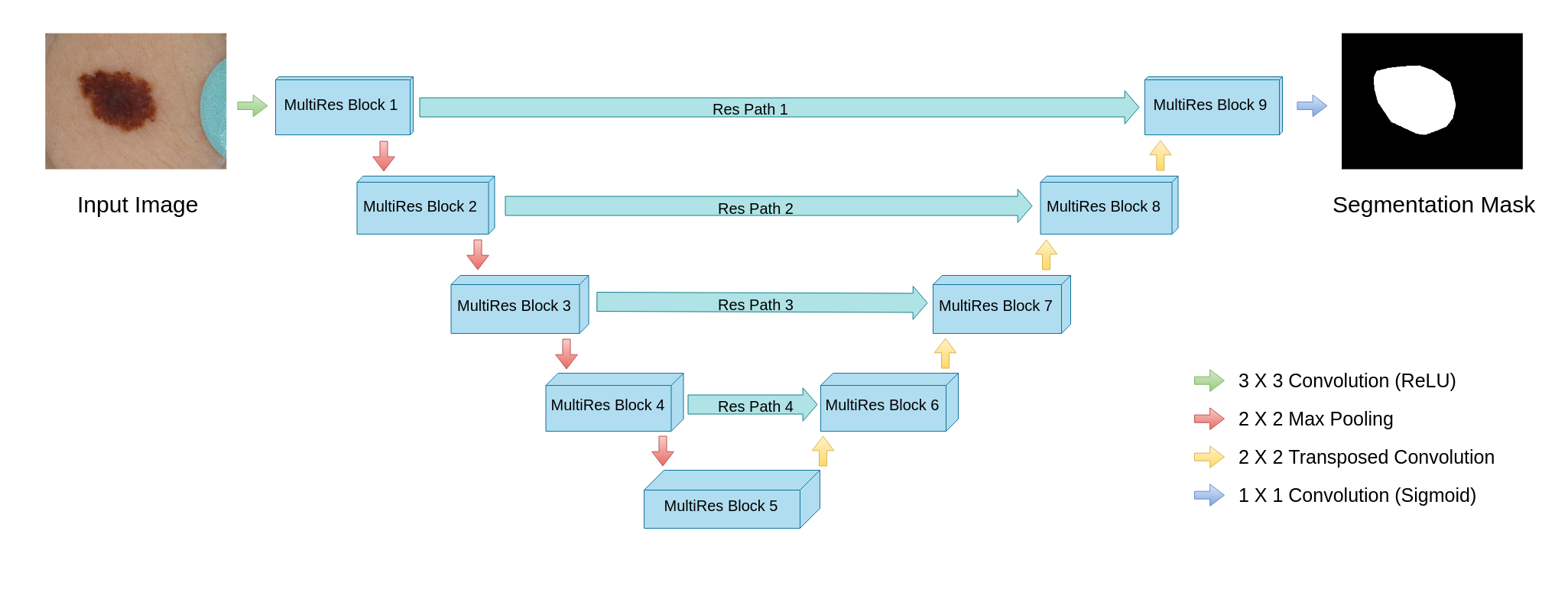}
        \caption{Network architecture}
        \label{fig:my_label2}
     \end{subfigure} 
 
     \begin{subfigure}[b]{0.45\textwidth}
          \includegraphics[width=\textwidth]{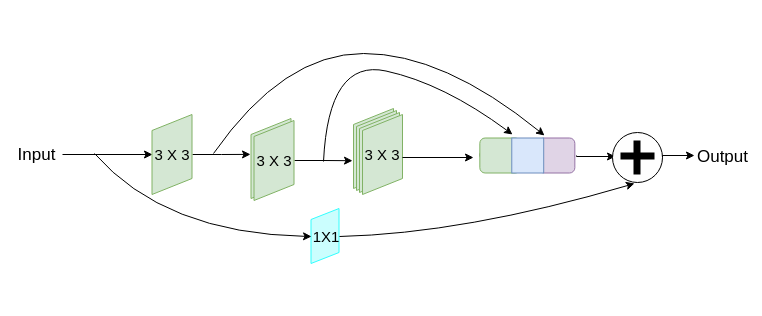}
        \caption{MultiRes block}
        \label{fig:mblock}
    \end{subfigure}    
     \begin{subfigure}[b]{0.45\textwidth}
          \includegraphics[width=\textwidth]{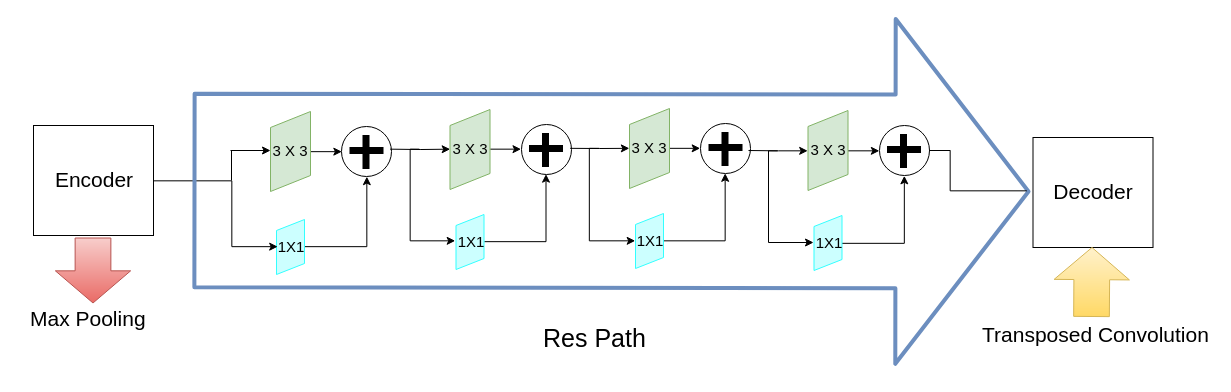}
        \caption{Res Path}
        \label{fig:respath}
    \end{subfigure}
    \caption{MultiResUNet architecture. Figure borrowed from \cite{ibtehaz2020multiresunet}.}
        \label{fig:multiresunet}
 \end{figure}
 
\item \textbf{Deep Supervision}\\
We have applied deep supervision on both U-Net and MultiResUNet models. Rather than only evaluating the output of the top layer's decoder, this model associates loss weights with the outputs of all five layers, in descending order from top to bottom. In order to achieve this, a simple 1 x 1 convolution is applied in all the layers after the up-Sampling and convolution operations. Figure \ref{fig:deepSN} presents Deeply Supervised U-Net and MultiResUNet models.
 
 

\begin{figure}[!htbp]
     \centering
     \begin{subfigure}[b]{0.8\textwidth}
        \includegraphics[width=\textwidth]{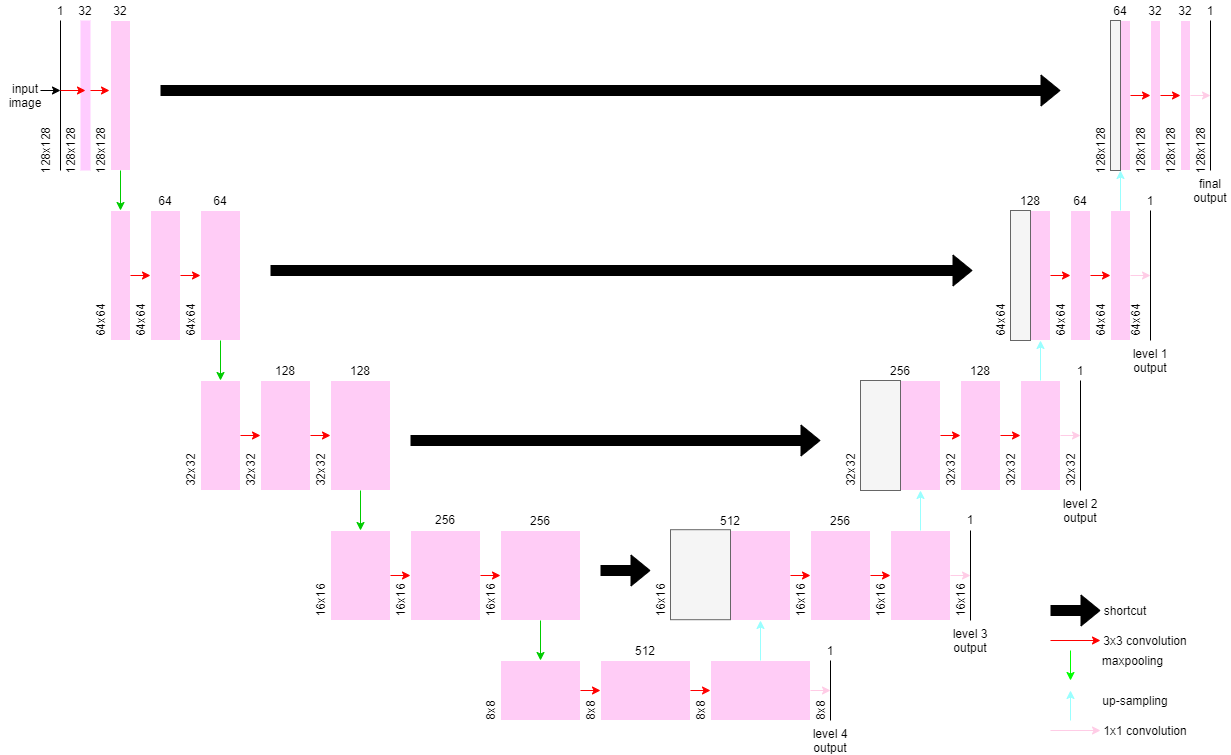}
     \caption{Deeply Supervised U-Net}
     \label{fig:DeepU}
     \end{subfigure}
 
  \begin{subfigure}[b]{0.8\textwidth}
     \centering
     \includegraphics[width=\textwidth]{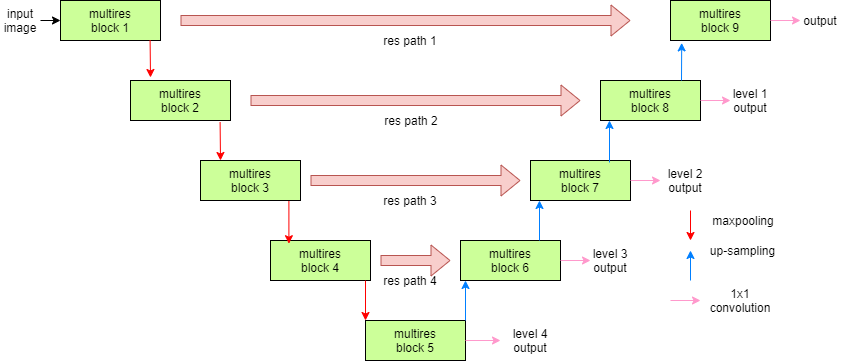}
     \caption{Deeply Supervised MultiResUNet}
     \label{fig:DeepM}
    \end{subfigure}
    
     \caption{Deeply Supervised Networks}
     \label{fig:deepSN}
     \end{figure}

\end{enumerate}

\subsection{The Baseline Model}

We chose U-Net \cite{ronneberger2015u} as our baseline model and applied deep supervision on U-Net as well as MultiResUNet in our experiments. Since the validation set of the original dataset was used for testing, a 5-fold cross validation was implemented. For performance evaluation, the metric called dice co-efficient was calculated. The value of dice co-efficient is always in the range between 0 and 1 (inclusive) 
and is calculated using the following formula:

\[ \mathrm{dice~co\mhyphen efficient} = \frac{2 * (|x \cap y|)}{|x| + |y|} \]

Here, $x$ and $y$ represent the tumor areas of the ground truth and the prediction, respectively. Following the literature \cite{kamal2018lung}, the dice co-efficient is computed as follows: (i) For True-Negatives (i.e., there is no tumor, and the processing algorithm correctly detected that), the dice co-efficient would be 1; and (ii) For False-Positives (i.e., there is no tumor, but the processing algorithm mistakenly segmented the tumor), the dice co-efficient would be 0.



\subsection{Coding and Environment}
Our experiments have been conducted in a desktop computer having Intel core i7-7700 processor (3.6 GHz, 8 MB cache) CPU, 16 GB RAM, and NVIDIA TITAN XP (12 GB, 1582 MHz) GPU. We have used Python3 programming language \cite{van2007python}. For implementing the models, we used Keras \cite{chollet2015keras} with Tensorflow backend \cite{abadi2016tensorflow}.

\section{Results}
\subsection{Model Parameters}
We have assessed the results by varying different parameters in our model - optimization algorithm, learning rate and decay of the network. This led to various dice co-efficients (training accuracy in this case) as reported in Table \ref{tab:table3}.

\begin{table}[!htbp]
    \centering
    \caption{Dice Co-efficients for different optimizers, learning rates and decay for UNet and MultiResUNet models}\label{tab:table3}
    \begin{tabular}{|c|c|c|c|c|}
        \hline
         \textbf{Model} & \textbf{Optimization} & \textbf{Learning Rate} & \textbf{Decay} & \textbf{Dice Co-efficient}\\
         &\textbf{Algorithm}&&&\textbf{(Training}\\
         &&&& \textbf{Accuracy)}\\
         \hline
         U-Net & Adam & 0.1 & 0.1/150 & 0.7218\\
         \hline
         U-Net & Adam & 0.01 & 0.01/150 & 0.7636\\
         \hline
         U-Net & Adam & 0.001 & 0.001/150 & 0.7591\\
         \hline
         U-Net & SGD & 0.1 & 0.1/150 & 0.5780\\
         \hline
         U-Net & SGD & 0.01 & 0.01/150 & 0.4525\\
         \hline
         MultiResUNet & Adam & 0.1 & 0.1/150 & 0.7437\\
         \hline
         MultiResUNet & Adam & 0.01 & 0.01/150 & 0.7533\\
         \hline
         MultiResUNet & Adam & 0.001 & 0.001/150 & 0.7435\\
         \hline
         MultiResUNet & SGD & 0.5 & 0.5/150 & 0.6556\\
         \hline
         MultiResUNet & SGD & 0.1 & 0.1/150 & 0.5916\\
         \hline
    \end{tabular}

\end{table}

It can be observed from Table \ref{tab:table3} that Adam has performed significantly better than Stochastic Gradient Descent (SGD) for both U-Net and MultiResUNet. Besides, for both the models, a learning rate of 0.01 has outperformed any other learning rate. We can infer this from the table since with an increase of the learning rate from 0.001 to 0.01, the dice co-efficient increases. However, it falls once the learning rate is raised to 0.1. 

Once we have fixed on the Adam optimizer (as the optimization algorithm) and a learning rate of 0.01, we now focus on deep supervision. Performance of deep supervision on U-Net and MultiResUNet models involves consideration of loss weights associated with the outputs from each level.  Table \ref{tab:table4} presents the dice co-efficient achieved during the training for various loss weights in different layers.

\begin{table}[!htbp]
    \centering
    \caption{Dice Co-efficients for deep supervision for various loss weights in different layers}\label{tab:table4}
    \begin{tabular}{|c|c|c|c|c|c|c|}
        \hline
         \textbf{Model} & \textbf{Loss} & \textbf{Loss} & \textbf{Loss} & \textbf{Loss} & \textbf{Loss} & \textbf{Dice}\\
         &\textbf{Weight} &\textbf{Weight} &\textbf{Weight} &\textbf{Weight} &\textbf{Weight} & \textbf{Coefficient}\\
         & \textbf{(Final} & \textbf{(Level} & \textbf{(Level} & \textbf{(Level} & \textbf{(Level} & \textbf{(Training}\\
         &\textbf{layer)}& \textbf{1)}&\textbf{2)}&\textbf{3)}&\textbf{4)}&\textbf{Accuracy)}\\
         \hline
         Deeply & 1.00 & 0.8 & 0.6 & 0.4 & 0.2 & 0.7574\\
         Supervised &&&&&&\\
         UNet &&&&&&\\
         \hline
         Deeply & 1.00 & 0.7 & 0.5 & 0.3 & 0.1 & 0.7582\\
         Supervised &&&&&&\\
         UNet &&&&&&\\
         \hline
         Deeply & 1.00 & 0.8 & 0.6 & 0.4 & 0.2 & 0.7482\\
         Supervised &&&&&&\\
         MultiResUNet &&&&&&\\
         \hline
         Deeply & 1.00 & 0.7 & 0.5 & 0.3 & 0.1 & 0.7421\\
         Supervised &&&&&&\\
         MultiResUNet &&&&&&\\
         \hline
    \end{tabular}

\end{table}

\subsection{Performance on the Test set}

We used our test set containing data from 40 subjects for inferring how well the trained models perform and which ones perform well when supplied with new instances. We have applied various threshold values and numbers of rotations during test time augmentation. 

\subsubsection{Tumor Detection}

Since the predicted probability of being tumor pixels is between 0 and 1, we have binarized it based on certain threshold values. For example, if the probability of a pixel for being a tumor is $x$ and the threshold is $t$ then for all $x > t$, our model predicts it to be a tumor, whereas for all $x \leq t$, the model does not recognize it as a tumor pixel. Table \ref{tab:table5} gives the true positives, false positives, true negatives, false negatives and F1 score for all the model settings.

\begin{table}[!htbp]
\begin{adjustwidth}{-.5in}{-.5in}  
        \begin{center}
    \small
    \centering
    \caption{All quantitative results related to tumor detection for different numbers of rotations and thresholds using the test set.}
    \label{tab:table5}
    \begin{tabular}{|c|c|c|c|c|c|c|c|}
        \hline
         \textbf{Model} & \textbf{Number}& \textbf{Thres}  & \textbf{True} & \textbf{False} & \textbf{True} & \textbf{False} & \textbf{F1}\\
         &\textbf{of}&\textbf{-hold} & \textbf{Positive} & \textbf{Positive} & \textbf{Negative} & \textbf{Negative} & \textbf{Score}\\
         &\textbf{Rotations}& &&&&&\\
         &&&&&&&\\
         \hline
         U-Net & 20 & 0.4 &104&130&3502&744&0.1922\\
         \hline
         U-Net & 50 & 0.5 &53&58&3574&795&0.1105\\
         \hline
         MultiResUNet & 20 & 0.4  &576&215&3417&272&0.7028\\
         \hline
         MultiResUNet & 50 & 0.5  & 447&115&3517&401&0.634\\
         \hline
         Deeply Supervised & 20 & 0.4 & 632&520&3112&216&0.6315\\
         U-Net &&&&&&&\\
         \hline
         Deeply Supervised & 50 & 0.5  &501&210&3422&347&0.6422\\
         U-Net &&& &&&&\\
         \hline
         Deeply Supervised & 20 & 0.4  &600&268&3364&248&0.6990\\
         MultiResUNet &&& &&&&\\
         \hline
         Deeply Supervised & 50 & 0.5  &505&123&3509&343&0.6838\\
         MultiResUNet &&& &&&&\\
         \hline
    \end{tabular}
    
    \end{center}
    \end{adjustwidth}
\end{table}

\subsubsection{Tumor Segmentation}
  Table \ref{tab:table6} reports the dice co-efficients for the test set with various numbers of rotations applied during test time augmentation. Moreover, in addition to performing standard binarization using a threshold of 0.5, we also experimented with different threshold values. Evidently, deep supervision has been able to improve the performance for both U-Net and MultiResUNet and the latter has performed better. More specifically, deeply supervised MultiResUNet model (with 50 rotations and 0.5 threshold) is the winner with a dice co-efficient of 0.8472. 

\begin{table}[!htbp]
\begin{adjustwidth}{-.5in}{-.5in}  
        \begin{center}
    
    \centering
    \caption{Dice co-efficients for different numbers of rotation and thresholds using the test set.}
    \label{tab:table6}
    \begin{tabular}{|c|c|c|c|}
        \hline
         \textbf{Model} & \textbf{Number of}& \textbf{Threshold} & \textbf{Dice} \\
         &\textbf{Rotations}&&\textbf{Co-efficient}\\
         &&&\textbf{(Test Accuracy)}\\
         \hline
         U-Net & 20 & 0.4 & 0.7882\\
         \hline
         U-Net & 50 & 0.4 & 0.7890\\
         \hline
         U-Net & 20 & 0.5 & 0.7984\\
         \hline
         U-Net & 50 & 0.5 & 0.8003\\
         \hline
         MultiResUNet & 20 & 0.4 & 0.8327\\
         \hline
         MultiResUNet & 50 & 0.4 & 0.8363\\
         \hline
         MultiResUNet & 20 & 0.5 & 0.8373\\
         \hline
         MultiResUNet & 50 & 0.5 & 0.8382\\
         \hline
         Deeply Supervised & 20 & 0.4 & 0.7645\\
         U-Net &&& \\
         \hline
         Deeply Supervised & 50 & 0.4 & 0.7858\\
         U-Net &&&\\
         \hline
         Deeply Supervised & 20 & 0.5 & 0.8158\\
         U-Net &&& \\
         \hline
         Deeply Supervised & 50 & 0.5 & 0.8224\\
         U-Net &&&\\
         \hline
         Deeply Supervised & 20 & 0.4 & 0.8294\\
         MultiResUNet &&& \\
         \hline
         Deeply Supervised & 50 & 0.4 & 0.8324\\
         MultiResUNet &&& \\
         \hline
         Deeply Supervised & 20 & 0.5 & 0.8434\\
         MultiResUNet &&& \\
         \hline
         Deeply Supervised & 50 & 0.5 & 0.8472\\
         MultiResUNet &&& \\
         \hline
    \end{tabular}
   
    \end{center}
    \end{adjustwidth}
\end{table}



\begin{figure}[!htbp]
    \centering
    \subfloat{{\includegraphics[scale=0.5]{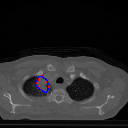} }}%
    \qquad
    \subfloat{{\includegraphics[scale=0.5]{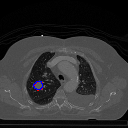} }}%
    \qquad
    \subfloat{{\includegraphics[scale=0.5]{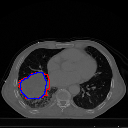} }}%
    \qquad
    \subfloat{{\includegraphics[scale=0.5]{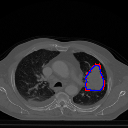} }}%

    \centering
     \subfloat{{\includegraphics[scale=0.5]{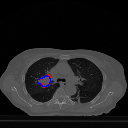} }}%
    \qquad
     \subfloat{{\includegraphics[scale=0.5]{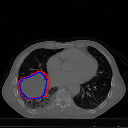} }}%
     \qquad
     \subfloat{{\includegraphics[scale=0.5]{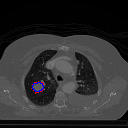} }}%
     \qquad
     \subfloat{{\includegraphics[scale=0.5]{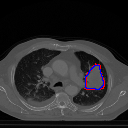} }}%
    \caption{Strength of our model. The \textcolor{red}{ground truth (in red)} and \textcolor{blue}{prediction (in blue)} for Deeply Supervised MultiResUNet model.}%
    \label{fig:example}%
\end{figure}

A qualitative analysis of our results reveals the strengths and weaknesses of our proposed model. In most cases, regardless of the size of the ground truth, our model predicts the tumor shapes very well. This is evident from Figure \ref{fig:example} where the ground truth and the prediction (by Deeply Supervised MultiResUNet) are shown in red and blue respectively. Although the tumors are in various arbitrary locations within the lung and appear in diverse sizes, the red and blue margins appear to coincide almost perfectly.

\begin{figure}%
    \centering
    \subfloat[\centering Prediction with MultiResUNet (Image 1)]{{\includegraphics[width=3cm]{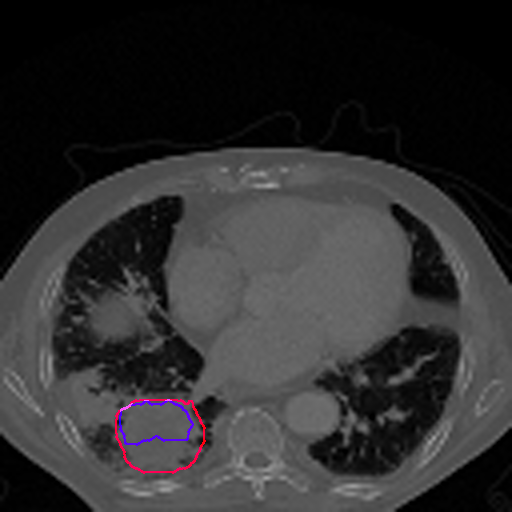} }}%
    \qquad
    \subfloat[\centering Prediction with Deeply Supervised MultiResUNet (Image 1)]{{\includegraphics[width=3cm]{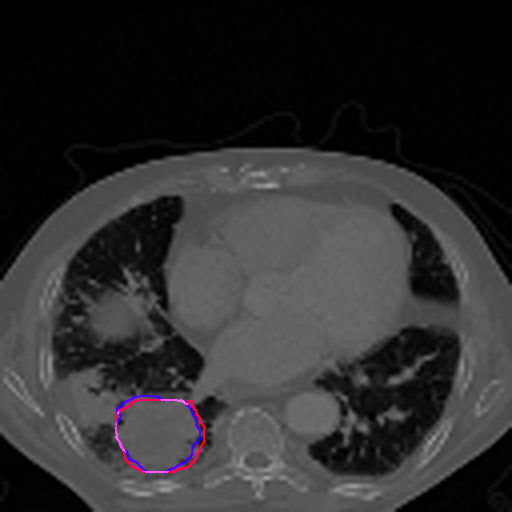} }}%

    \subfloat[\centering Prediction with MultiResUNet (Image 2)]{{\includegraphics[width=3cm]{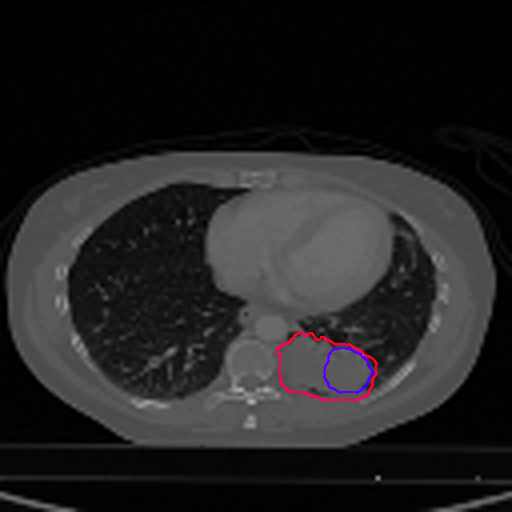} }}%
    \qquad
    \subfloat[\centering Prediction with Deeply Supervised MultiResUNet (Image 2)]{{\includegraphics[width=3cm]{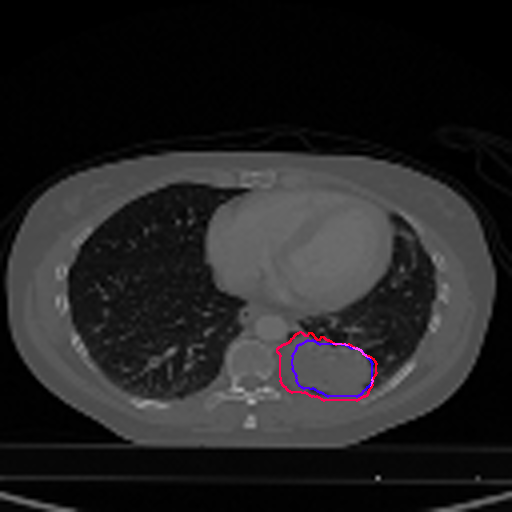} }}%
    
    \caption{Deeply Supervised MultiResUNet outperforms MultiResUNet significantly in terms of capturing the range of the tumor. The ground truth is shown in red whereas the prediction is given in blue.}%
    \label{fig:compareDS}%
\end{figure}

Figure \ref{fig:compareDS} shows a comparison between predictions by the MultiResUNet model and Deeply supervised MultiResUNet model. It can be seen that the latter can delineate the tumor edges more accurately than the former. The ground truth (shown in red) and the prediction (shown in blue) have a more consistent alignment with each other in Deeply Supervised MultiResUNet's prediction.

\begin{figure}[!htbp]
    \centering
    \subfloat{{\includegraphics[scale=0.15]{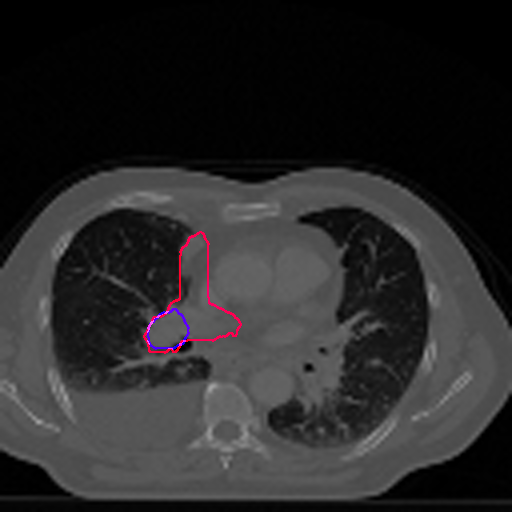} }}%
    \qquad
    \subfloat{{\includegraphics[scale=0.15]{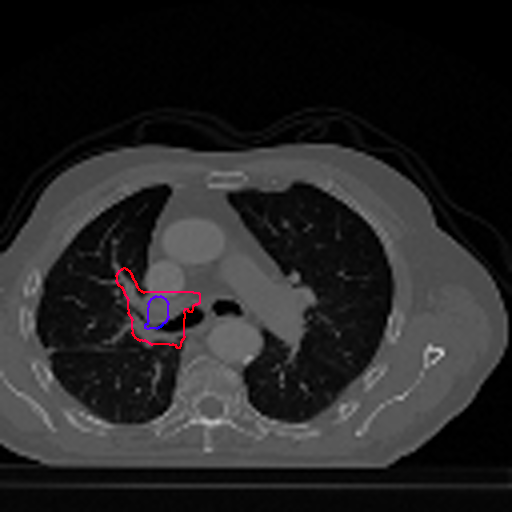} }}%
    \qquad
    \subfloat{{\includegraphics[scale=0.15]{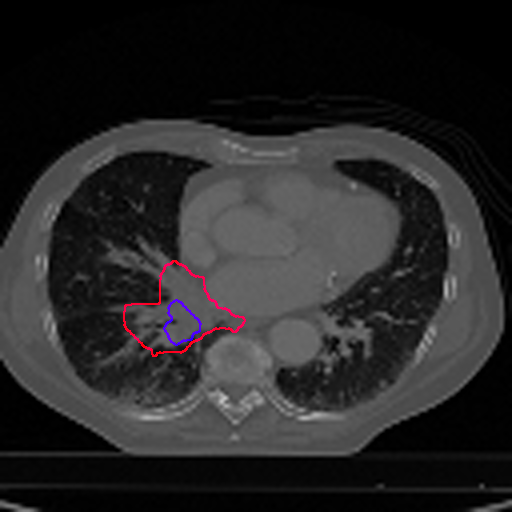} }}%
    
    \caption{Limitation of our model. Tumors with tremendously erratic boundaries are predicted imperfectly. The ground truth is shown in red whereas the prediction is given in blue.}%
    \label{fig:limitation}%
\end{figure}

However, in a few rogue cases, where the shapes of the ground truth are exceedingly uncoordinated, perhaps with fissures in the middle or with erratic outlines, it is seen that the predicted tumor regions are prone to imperfections. Although for these cases, our model can predict the existence of tumors, the shapes can often appear to have disfigurements. This is demonstrated in Figure \ref{fig:limitation}. We plan on addressing this issue in our future work.

\subsection{Comparison with other models}

A comparative analysis involving other state-of-the-art models shows that our model outperforms all of them by a significant margin. This is presented in Table \ref{tab:compare} which shows the mean dice co-efficient obtained using our model and several others. 

At this point a brief discussion on the (unusual) dice calculation of Hossain et al. \cite{hossain2019pipeline} is in order, which differs from the calculation followed in \cite{kamal2018lung} (and this work). 
The test set in \cite{hossain2019pipeline} contained 4478 slices from 40 subjects. Their frontend binary classifier produced 1158 false negatives out of those but this huge number of misclassifications did not affect the dice score, as only the slices with tumor pixels in the ground truth were considered during their dice score calculation. Although they mentioned that the number of false positives was reduced by approximately 50\% through segmentation and post-processing, the paper makes no attempt to address what happened to the true positives reported by the binary classifier after passing them through the segmentation model and therefore, conveniently ignores the possibility of a reduction in the number of true positives which is a major issue.

Let us now analyze the the quantitative results reported in \cite{kamal2018lung}. The highest dice co-efficient achieved by them is 0.7228. For this result, the false positives and false negatives are 321 and 331 in number, respectively \cite{kamal2018lung}. In our case, the best dice co-efficient (of 0.8472) is achieved by the Deeply Supervised MultiResUnet which gives 123 false positives (much less than \cite{kamal2018lung}) and 343 false negatives (close to \cite{kamal2018lung}). Notably, we have already achieved far less false negatives for Deeply Supervised MultiResUnet with 20 rotations and 0.4 threshold (248 in number) at the cost of a slightly worse dice score (i.e., 0.8294) and false positives (268 in number)- both far better than that of \cite{kamal2018lung}. However, we still plan on exploring ways to reduce both false positives and negatives further while maintaining a high dice co-efficient. 

\begin{table}[!htbp]
    \centering
    \caption{Dice co-efficients obtained using different models in lung tumor segmentation task. Values for the other models have been taken from the respective papers (\cite{hossain2019pipeline} and \cite{kamal2018lung})}
    \label{tab:compare}
    \begin{tabular}{|c|c|}
        \hline
         \textbf{Model} & \textbf{Mean Dice Co-efficient}\\
         \hline
         2D-LungNet \cite{anthimopoulos2018semantic} & 0.6267\\
         \hline
         3D-LungNet \cite{hossain2019pipeline} & 0.6577\\
         \hline
         3D-DenseUNet \cite{kamal2018lung}, \cite{kolavrik2019optimized} & 0.6884\\
         \hline 
         Recurrent 3D-DenseUNet \cite{kamal2018lung} & 0.7228\\
         \hline
         \textbf{Deeply Supervised} & \textbf{0.8472}\\
         \textbf{MultiResUNet} &\\
         \hline
    \end{tabular}
    
\end{table}

\subsection{Ablation Study}
We conducted an ablation study to analyze the performance of our overall pipeline. In particular, we check what happens if wavelet transforms are not applied to the dataset in the preprocessing step. We preprocessed the images without wavelet transforms and only kept the neighboring slices along with the original image. This resulted in a 3-channel input instead of a 5-channel one wherewith we then trained the Deeply Supervised MultiResUNet with loss weights of [1.00,0.8,0.6,0.4,0.2] while keeping all other parameters same as before. This same model when trained on the dataset containing wavelet transforms gives a dice co-efficient of 0.8294 on the test set (after applying 20 rotations and with a threshold of 0.4). On the other hand, the model trained on the dataset without wavelet transforms gives a dice co-efficient of 0.7649 with the same setting. 

Moreover, we performed a qualitative analysis on the results produced from the two cases. We observed two things that happen in the absence of wavelet transforms -
\begin{enumerate}
    \item The predicted tumors seem to have frequent outliers
    \item In most cases, the predicted tumor seems to be enlarged and the edges of the tumors are not properly detected
\end{enumerate}

\begin{figure}%
    \centering
    \subfloat[\centering Presence of outlier without wavelet transform]{{\includegraphics[width=3cm]{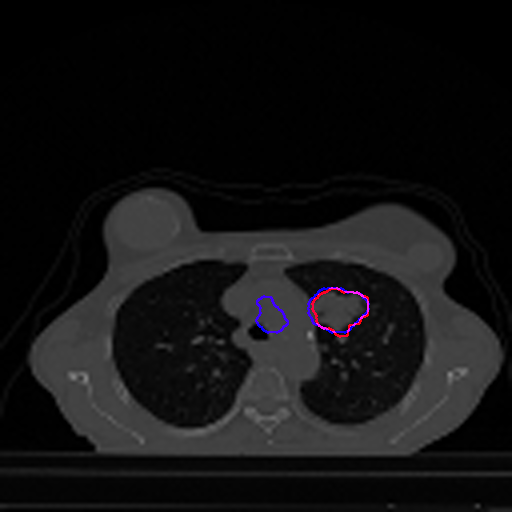} }}%
    \qquad
    \subfloat[\centering Absence of outlier with wavelet transform]{{\includegraphics[width=3cm]{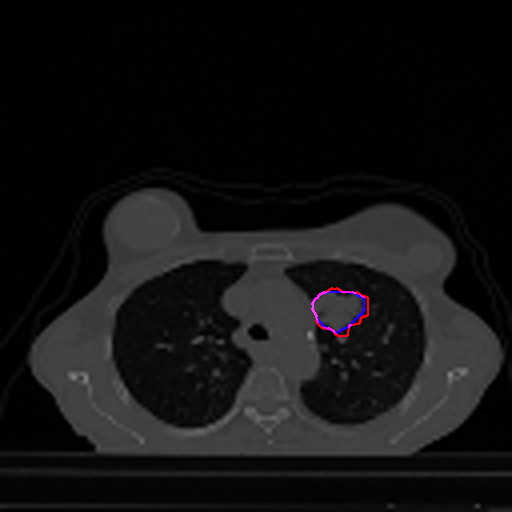} }}%

    \subfloat[\centering Enlarged prediction without wavelet transform]{{\includegraphics[width=3cm]{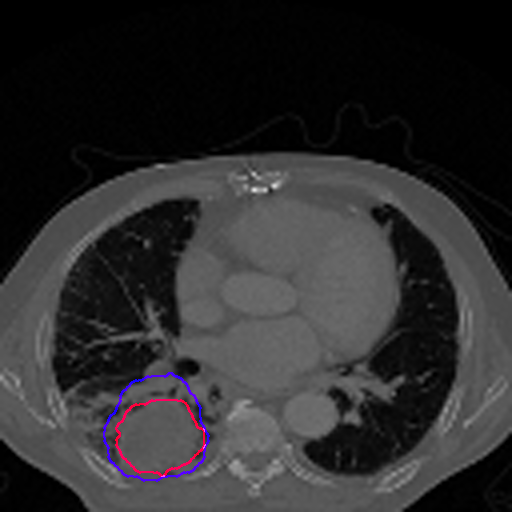} }}%
    \qquad
    \subfloat[\centering Better edge prediction with wavelet transform]{{\includegraphics[width=3cm]{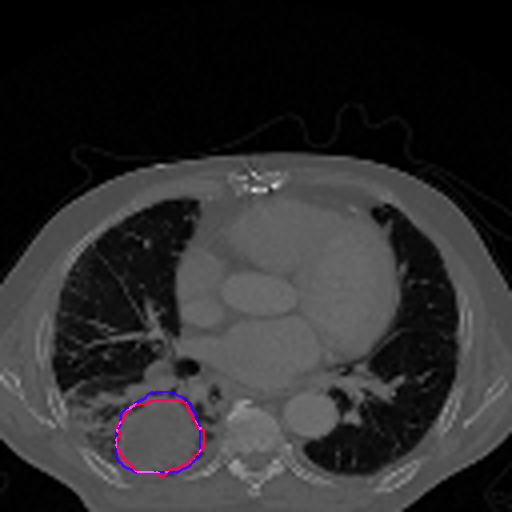} }}%
    
    \caption{Application of wavelet transform reduces the frequency of outliers and avoids an inflated predicted tumor size. The ground truth is shown in red whereas the prediction is given in blue.}%
    \label{fig:compareWV}%
\end{figure}

These observations have been demonstrated in Figure \ref{fig:compareWV}.

\section{Discussions}

MultiResUNet outperforms U-Net in segmentation of lung tumor from CT slices and deeply supervised MultiResUNet performs even better. In particular, due to deep supervision the best MultiResUNet model registers an improvement from 0.8382 to 0.8472 in terms of dice co-efficient. Since deep supervision considers the outputs from the lower levels of the decoder, the reconstruction phase of the model begins at a very deep level. This positively affects the final dice co-efficient calculated on the predictions. Besides, we believe, wavelet transforms combined with using neighboring CT slices during input instance preparation have allowed our model to analyze tumor textures more accurately.

Table \ref{tab:table5} demonstrates various quantitative results for tumor detection. A crucial observation is that, for every model, application of deep supervision reduces the number of false negatives. Reducing false negatives is crucial as failing to identify a sick patient with tumor has serious detrimental effects. Tables \ref{tab:table5} and \ref{tab:table6} illustrate another pattern. The models which give a higher F1 score also yield better dice co-efficients although the margin may not be very high.

Table \ref{tab:table6} shows a pattern of increase in dice co-efficients as we increase the thresholds and number of rotations. For each segmentation model, keeping the threshold constant, with the increase of rotations, the dice co-efficient increases as well.
The rationality behind this is that, when the number of rotations is low, it is difficult to filter out the false positive pixels as opposed to a greater number of rotations where such pixels receive a very low probability of being tumors. When rotations are more, the ability to confidently claim that a certain pixel will be part of the tumor is more. Having a higher threshold has a similar effect on the predictions. This is because a pixel is only considered a part of the tumor if the predicted probability is quite high. Otherwise, it is discarded. This effectively filters out the false positive pixels. 

Another observation from Table \ref{tab:table6} is that, with the increase in threshold, the performance of deep supervision improves as well. This can be seen in the case of both U-Net and MultiResUnet. Keeping the number of rotations constant, an increase in threshold results in a improved dice co-efficient for deep supervision. We observed that the deeply supervised versions of the model almost consistently had probability scores higher than 0.5 in the tumor regions. So, when we consider threshold values less than 0.5 (i.e., 0.4), very little additional actual tumor pixels were considered compared to the background pixels having scores at the range of 0.4-0.5. As a result, it slightly dips the performance. On the other hand, for models without deep supervision, there exists a bigger ratio of tumor pixels in the 0.4-0.5 range, thus the performance is better compared to the deep supervised models. Our takeaway point was that, apparently, for deep supervised models, the separation between background and foreground is a bit better. 

Figure \ref{fig:compareWV} shows the occurrence of outliers in the absence of wavelet transforms and an inability to accurately determine tumor edges in this case. The intuition concerning the former observation is that without wavelet transforms, texture information is relatively lost and the model perceives similarly shaped structures to be tumors as well. The second observation is also related to the fact that wavelet transforms can capture the texture better. In absence of it, the model cannot predict or justifiably limit the extent of the area that the tumor boundary should encompass. This results in inflated tumor areas and inaccurate tumor sizes.

The mechanism that we have proposed in this paper is unique, resulting in a procedure that returns an excellent dice co-efficient value of 0.8472. The technique of combining neighboring slices for more detail and applying wavelet transforms for texture analysis have never been implemented in this way before. The first and second approximations of the original slice, obtained from two-dimensional discrete wavelet transform, have allowed the model to peruse details that are lost, concluding with a finer interpretation of the input. We have also implemented Deeply Supervised MultiResUNet, assigning priorities to the outputs of lower layers of the model. Apart from these, application of Test Time Augmentation has assisted the network to execute better when supplied with new instances. 

\section{Conclusion}
The battle against lung cancer has been going on for a very long time. To this end, a pivotal field of research has emerged and is gaining traction with the help of deep learning in the recent years. The proposed methods in our paper performed significantly better than other existing approaches in this attempt towards detecting lung tumors with CT Scan images. The strategy that we undertook included two-dimensional discrete wavelet transform during the preprocessing step and experimenting with multiple models that led to splendid results - in terms of dice co-efficient - especially in the case of Deeply Supervised MultiResUNet model which ensured more accurate delineation of faint boundaries. 


We believe that our approach can make the tasks of oncologists much easier and help detect tumors at an early stage. We have planned for further research regarding this particular topic that encompasses refining the pipeline in order to achieve even better results. For example, we plan to explore other preprocessing techniques apart from wavelet transforms thereby redesigning the pipeline. Moreover, we plan on experimenting with additional techniques that could improve the performance of the models that we used. One such interesting avenue to explore is to use attention mechanism with the MultiResUnet model.

\bibliographystyle{IEEEtran}
\bibliography{references}

\end{document}